\documentclass[preprint,showpacs,showkeys,aps,prd,amsfonts,eqsecnum,nofootinbib]{revtex4}

\usepackage{graphicx,epsfig}
\usepackage[dvips]{color}

\newcommand{\beq}{\begin{equation}}
\newcommand{\eeq}{\end{equation}}
\newcommand{\bea}{\begin{eqnarray}}
\newcommand{\eea}{\end{eqnarray}}
\newcommand{\rf}[1]{(\ref{#1})}

\begin{document}
\preprint{CUMQ/HEP 141}
%
%
\title{\Large THE CASIMIR FORCE IN A LORENTZ VIOLATING THEORY}

\author{Mariana Frank}\email[]{mfrank@vax2.concordia.ca}
\author{Ismail Turan}\email[]{ituran@physics.concordia.ca}

\affiliation{
Department of Physics, Concordia University, 7141 Sherbrooke
Street West, Montreal, Quebec, CANADA H4B 1R6}
\date{\today}

\begin{abstract}

We study the effects of the minimal extension of the standard model including Lorentz violation on the  Casimir force between two parallel conducting plates in vacuum. We provide explicit solutions for the electromagnetic field using scalar field analogy, for both the cases in which the Lorentz violating terms come from the CPT-even or CPT-odd terms. We also calculate the effects of the Lorentz violating terms for a fermion field between two parallel conducting plates and analyze the modifications of the Casimir force due to the modifications of the Dirac equation. In all cases under consideration, the standard formulas for the Casimir force are modified by either multiplicative or additive correction factors,   the latter case exhibiting different dependence on the distance between the plates. 
\end{abstract}
\maketitle
\section{Introduction}\label{sec:intro}
The search for a theory beyond the minimal $SU(3)_C \times SU(2)_L \times U(1)_Y$ standard model (SM) 
is motivated by the fact that,
although phenomenologically successful, the SM suffers from some theoretical inconsistencies, 
and from 
some long standing unresolved problems. More general scenarios exist, in which  SM is viewed as
 a low-energy limit of more fundamental theory, which should be able to
 provide a quantum description of gravitation. 

An interesting alternative at the Planck scale is 
the possibility that the new physics scenario 
involves a violation of Lorentz symmetry.
In particular, in the context of string theories,  it has been shown  
that spontaneous breaking of Lorentz symmetry 
may occur
with Lorentz-covariant dynamics
\cite{kps}.
In these theories, interactions are
triggered by
nonzero expectation values for Lorentz tensors, because  
 spontaneous breaking of 
the higher-dimensional Lorentz invariance is expected 
in any realistic Lorentz-covariant fundamental theory 
involving more than four space-time dimensions.
If the breaking extends
into the four macroscopic spacetime dimensions,
 the Lorentz symmetry violation could appear 
at the level of the SM. 
Because the breaking is spontaneous, 
Lorentz symmetry remains a property
of the underlying fundamental theory. 
 Another important property of the spontaneous breaking
is that the theory remains invariant under
{\it observer}  Lorentz transformations,
i.e., under rotations and boosts of an observer's inertial frame.  
In addition to that, conventional quantization,
hermiticity, gauge invariance, power-counting renormalizability,
and the expected microcausality and positivity
of the energy are also maintained. Note also that there are other 
ways to break Lorentz symmetry. 
It can, for example, occur dynamically in quantum field theories 
\cite{Andrianov:1994qv}, or via a CPT anomaly in compact spaces 
\cite{Klinkhamer:1999zh}.

Such a framework  for examining the effects 
of spontaneous Lorentz breaking 
in the context of a low-energy effective theory, known as the Standard-Model Extension (SME), has been developed 
explicitly by Colladay and Kosteleck\'y first without gravity \cite{cksm} 
and later by Kostelecky with gravity \cite{akgrav}.  Gravitationally coupled SME is described in 
non-Minkowski spacetimes and leads to spacetime-dependent
 coefficients; and recently, the pure gravity sector of the model has been studied \cite{akgrav}. In this paper, 
we confine the framework into the minimal SME without gravity.\footnote{From now on, we drop the phrase ``minimal" 
and use only ``SME" to refer to the minimal version.} General reviews of the model and the status of the recent experimental investigations can be found in Refs.~\cite{cpt04,reviews}. An extensive analysis of the implications of the model for electrostatics and magnetostatics in SME, as well as discussion of its QED sector can be found in \cite{Bailey:2004na, photonth1, photonth2}. 

The SME
\cite{cksm}
has been the focus of various experimental studies, 
including ones
with photons, from radiative corrections, photon splitting
and vacuum \v Cerenkov and synchrotron radiation \cite{photonexpt,photonth1,photonth2},
electrons \cite{eexpt,eexpt2,eexpt3},
protons and neutrons, including baryosynthesis and nucleosynthesis in \cite{ccexpt,spaceexpt,bnsyn},
CPT phenomena in mesons \cite{hadronexpt},
muons \cite{muexpt},
neutrinos \cite{nuexpt},
and Higgs bosons \cite{higgs}.
Though 
no evidence for Lorentz violation has yet been found,
only less than a half of the possible tests
involving light and ordinary matter 
(electrons, protons, and neutrons)
have been performed,
while some of the other sectors remain virtually unexplored.

Our goal here is to provide additional theoretical predictions  regarding the quantum vacuum in this extended theory, in particular we concentrate on calculating the effects of the minimal SME on the Casimir force. 

Zero-point fluctuations in quantum fields give rise to macroscopically observable forces between material bodies,  the so-called Casimir forces \cite{Casimir}.  In general, the Casimir effect can be defined as the stress (force per unit area) on bounding surfaces when a quantum field is confined in a finite volume of space. The boundaries can be material media, interfaces between two phases of the vacuum, or topologies of space (such as in higher dimensional theories). The simplest test is measuring the force between two parallel conducting plates. Simply imagined, the vacuum is a sea of waves of energy of all possible length, while between the plates only waves whose wavelengths exactly fit between the plates are present. One calculates all of the zero-point energy between the plates, and the result (which is divergent) is regulated by subtracting the value of the energy when no boundaries are present \cite{Milton:2004ya}. Experiments testing the Casimir force are very precise tests of the predictions of the field theory.
Several experimental attempts have been pursued during various decades since its theoretical prediction, for
unambiguously verifying Casimir's formula \cite{Bordag}.     
Early searches have been successful only in a particular geometry, namely 
in a cavity constructed by a plane surface opposing a spherical one. 
Pioneering measurements by van Blokland and Overbeek \cite{Vanblokland} 
in such a configuration resulted in the observation of the associated Casimir 
force, and in its detailed comparison to the Lifshitz theory \cite{Lifshitz}, by  
taking into account finite conductivity effects.
More recently, these measurements have been performed using  
torsion balances \cite{Lamoreaux}, atomic force microscopes \cite{Mohideen}, 
and high precision capacitance bridges \cite{Chan}. The latter two 
experiments have reached 1\% precision, more precise determinations 
being limited by theoretical uncertainties.  Sparnaay \cite{Sparnaay} has attempted the first experiments with parallel plates configuration, while Bressi {\it et al} \cite{Bressi} have recently reported a measurement of the coefficient of the Casimir force between 
parallel conducting surfaces in the $0.5 - 3.0 \,\,\mu$m range with 
15 \% precision: $K_C=(1.22\pm 0.18 ) \times 10^{-27}$ N m$^2$.

In this work, we investigate the Casimir effect in the vacuum of  the SME. The Casimir force is a very sensitive measurement of the quantum fluctuations of the field at the macroscopic level. Since the stress between two parallel conducting plates separated by a distance of $1~ \mu$m is $K_C/(1\mu m)^4$, where $K_C$ is of order of ${\cal O}(10^{-27}$ N m$^2$),  and with the hope that future experiments will improve the existing precision, this measurement should be tested against any new theory. We  evaluate the stress between parallel plates, leaving evaluations of quantum fluctuations for dielectric media and different topologies for later discussions.

Our paper is organized as follows. In Section II we summarize the SME by giving only the relevant terms in the Lagrangian.  In the subsequent sections, we calculate the Casimir force  for  scalar fields with two different frequencies of propagation, according to the SME (Section III) in the case of both CPT-even and CPT-odd  Lorentz violation terms. We examine the correction for  the fermion field (Section IV) and conclude and summarize our results in (Section V), where we also discuss the possibility of observing the deviations from the measured Casimir force.

\section{The Lorentz Violating Model}
\label{model}

The minimal extension of the Standard Model (SME) as given in \cite{cksm, akgrav, cpt04, reviews}
contains all possible Lorentz-violating terms
that could arise from spontaneous symmetry breaking
at a fundamental level, 
but that preserve
SU(3) $\times$ SU(2)$ \times$ U(1) 
gauge invariance 
and power-counting renormalizability.
All terms that are even or odd under CPT
are explicitly given in Ref.\ \cite{cksm} and will not be repeated here. We rather concentrate on 
the relevant part of the theory, namely the extended quantum electrodynamics drived from SME. 

The general form of a Lorentz-violating term
involves 
a part constructed from the basic fields in the standard model, whose strength is given by a coupling coefficient.
This imposes various limitations 
on the possible structures of both the operators and the couplings.
Taken together,
these requirements place significant constraints
on the form of allowed terms in the SME. 
Taken from the full SME that contains all known particles, the Lagrangian involving the Dirac 
field $\psi$ of the electron and the electromagnetic field $F^{\mu\nu}$ can be written as
\begin{eqnarray}
{\cal L}^{QED}&=&\frac{i}{2}\bar{\psi}\Gamma_{\nu}D^\nu \psi -\frac{1}{2}\bar{\psi}M\psi + {\rm H.c.} \nonumber\\
&&- \frac 1 4 F^{\mu\nu}F^{\mu\nu}
-\frac 1 4 (k_F)_{\kappa\lambda\mu\nu} F^{\kappa\lambda}F^{\mu\nu}
+ \frac 1 2 (k_{AF})^\kappa \epsilon_{\kappa\lambda\mu\nu} A^\lambda F^{\mu\nu}\,,
\label{EQEDlag}
\end{eqnarray}
where H.c. is the hermitian conjugate of the fermion part. The Lorentz violating parameters in the photon sector are included in the components of the tensors $(k_F)_{\kappa\lambda\mu\nu}$ (CPT-even) and $(k_{AF})^\kappa$ (CPT-odd). For the electron, the Lorentz violating parameters are encoded in $\Gamma_{\nu}$ and $M$ as
\begin{eqnarray}
\Gamma_\nu &=& \gamma_\nu + (c_{\mu\nu} + d_{\mu\nu}\gamma_5)\gamma^\mu + e_\nu+if_\nu\gamma_5+\frac{1}{2}g_{\kappa\mu\nu}\sigma^{\kappa\mu}\,,\nonumber\\
M &=& m + (a_\mu +b_\mu \gamma_5)\gamma^\mu+\frac{1}{2}H_{\mu\nu}\sigma^{\mu\nu},
\end{eqnarray}
where the parameters in $\Gamma_\nu$ are dimensionless while the ones in M have dimension of mass. Similarly, in the photon sector, $(k_F)_{\kappa\lambda\mu\nu}$ is dimensionless and $(k_{AF})^\kappa$ has dimension of mass. 
The parameters $e_\nu, f_\nu$ and $g_{\kappa\mu\nu}$ which are not extractable directly from SME are taken to be zero or very suppressed due to the renormalizibilty and gauge invariance requirements. 
$a_\mu$ can also be set to zero since it is not physical (it can be absorbed by using gauge transformations). 
The parameters $c_{\mu\nu}$ and $d_{\mu\nu}$ are traceless\footnote{Any non-zero trace would not contribute to Lorentz violation
and  can be absorbed by a conventional field normalization of the
 usual kinetic operator for the matter fields.} and $H_{\mu\nu}$ is antisymmetric. In the photon sector, $(k_F)_{\kappa\lambda\mu\nu}$
 has the symmetries of Riemann tensor plus a double traceless constraint, giving 19 independent components. Even though the components of $(k_{AF})^\kappa$ are strongly constrained  by cosmological observations, we allow them to be non-zero in our 
theoretical investigation of Casimir force. 

After presenting the terms in the Lagrangian of the extended QED, we are concerned with the modified QED vacuum. 
So, one can re-examine the basic features of electromagnetic radiation in this framework.  
Working with the standard parametrization of the plane wave with wave 4-vector $p^\alpha = (p^0, \vec p)$:
$F_{\mu \nu} (x)= F_{\mu \nu} (p) e^{-i p_\alpha x^\alpha}$, 
the equation determining the dispersion relation and the electric field $\vec E$ in the SME
is the modified Amp\`ere Law
\begin{equation}
M^{jk}E^k \equiv
\left[-\delta^{jk} p^2 - p^j p^k -2(k_F)^{j \beta \gamma k} p_\beta p_\gamma- 2 i (k_{AF})_\beta \epsilon^{j \beta \gamma k} p_\gamma\right]E^k = 0\ .
\label{ampere}
\end{equation}
The dispersion relation is obtained 
by requiring the vanishing of determinant of $M^{jk}$, and one obtains two modes for the radiation field \cite{cksm}:
\begin{equation}
p^0_{\pm} = (1 + \rho) |\vec p| \pm \sqrt{\sigma^2 {\vec p}^2 + \tau^2}
\quad ,
\label{disp}
\end{equation}
where
\begin{eqnarray}
\rho = -\frac 1 2{\tilde k}_\alpha^{~\alpha},
\quad
\sigma^2 =\frac 1 2({\tilde k}_{\alpha\beta})^2-\rho^2,
\quad
\tau = (k_{AF})^\mu\, {\hat p}_\mu
\label{rhsi}
\end{eqnarray}
with
\begin{eqnarray}
{\tilde k}^{\alpha\beta}&=&(k_F)^{\alpha\mu\beta\nu}{\hat {p}}_\mu {\hat {p}}_\nu,
\quad
{\hat {p}}^\mu = {p^\mu}/{|\vec p|}.
\end{eqnarray}

For vanishing $k_{AF}$ coefficients, and to leading order in the coefficients $k_F$ for Lorentz violation it becomes:
\begin{equation}
p^0_\pm=(1+\rho\pm\sigma)|\vec p|\ ,
\label{dispersion}
\end{equation}
Here ${\vec p}^{\,\,2}\rho$ and ${\vec p}^{\,\,2}\sigma$
are observer Lorentz scalars,
which implies $\rho$ and $\sigma$
are scalars under observer rotations. Note that in the leading order, 
${\hat {p}}_\mu=(1,{\vec p}/{|\vec p|})$ is assumed.

 The dispersion relation \rf{dispersion} has two solutions,
 corresponding to electric field values $\vec E_\pm$.
In conventional electrodynamics, 
the dispersion relation is $p^0=|\vec p|$
and thus the propagation is independent of the polarization.
However, 
in the present case
the propagation is goverened by two specific modes
$\vec E_\pm$,
with the general solution to \rf{ampere}
being any linear combination of the two, leading the phenomenon 
known as birefringence \cite{cksm}. This represents a fundamentally different description of photon fields propagating through the vacuum from conventional QED. The vacuum Lorentz breaking extension of electrodynamics resembles more electrodynamics in moving media \cite{cksm, Bailey:2004na}. Thus we expect changes in the vacuum energy, which impact  on the evaluation of the Casimir force.

\section{The scalar field}
\label{scalar}

Two main methods exist in the literature to calculate the vacuum energy of a field confined between  parallel plates. In the simplest and most transparent method, one  simply sums over all the modes of the fields in between the plates and then regularizes the result by subtracting the energy in the free space.  A more rigorous technique uses 
 consistent Green's functions and is presented in the following subsection.
Summing the modes  correctly produces
the finite, observable force starting from a divergent formal expression,
without any explicit subtractions, and is therefore of great utility
in practice. In what follows, we assume that the photon field can be approximated by two scalar fields of frequency $\omega_{\pm}=(1 + \rho) |{\vec p}| \pm \sqrt{(\sigma^2 {\vec p}^2 + \tau^2)}
$. Following normal procedure, we evaluate 
the correction to the usual Casimir energy by 
taking first $(k_{AF})^\kappa=0, (k_F)_{\kappa\lambda\mu\nu} \ne 0 $ (Case~I), 
then $(k_{AF})^\kappa \ne0, (k_F)_{\kappa\lambda\mu\nu} = 0 $ (Case~II).

 \subsection{Case I: $k_{AF}=0$}
\subsubsection{Summing of the Modes}
For simplicity,  the electromagnetic field between
parallel, uncharged, perfectly conducting plates in the SME due to the CPT-even Lorentz-violating parameter in the Lagrangian is seen as two massless scalar fields $\phi$
confined between two parallel
plates separated by a distance $D$.  In conventional QED, one accounts for the two polarization modes by simply multiplying the scalar result by 2. In SME, the two scalar fields will propagate with different frequencies, due to birefringence. 
 The scalar field satisfies Dirichlet boundary conditions on the plates,
that is: 
\begin{equation}
\phi(0)=\phi(D)=0.
\label{eq:bc}
\end{equation}
The Casimir force between the plates results from the zero-point energy 
per unit transverse area
\begin{equation}
u={1\over2}\sum \hbar\omega={1\over2}\sum_n \hbar(\omega_{n,1}+ \omega_{n,2})\,,
\end{equation}
where $\omega_{n,1}$ and $\omega_{n,2}$ represent the two zero-point energies associated with the surface modes, labeled by the positive integer n, \cite{Barton} of the electromagnetic  fields in the SME. The formalism bears some resemblance to the treatment of the force in a dispersive medium of effective ``dielectric" constant $\epsilon$ where $\epsilon_{\pm}=(1+\rho \pm \sigma)^2$. \footnote{This is a special case of the Lifschitz theory \cite{Lifshitz}, where the Casimir force is evaluated in a dielectric medium between two arbitrary parallel dielectrics, in the limit where the plates are perfect conductors.} The so-called surface modes are associated with the zero of the wave  number:
\begin{equation}
\kappa_{1,2}^2={\vec k}^2-\omega^2_{\pm}={\vec k}^2-\epsilon_{\pm} \omega^2\,,
\end{equation}
where we set $\hbar=c=1$, where $\vec k$ is  the transverse momentum.  
for the simple case where  $\omega_\pm \equiv p^0_\pm = (1 + \rho \pm \sigma) \omega$, $\omega$ being the frequency in the vacuum in the absence of dispersion. In this case (which corresponds to the conventional QED vacuum), the expression for the frequency of the field would be $\omega=|\vec k|$ for both surface modes; while in SME the frequencies of the two surface modes are $\displaystyle  \omega_{1,2}=  \frac{\omega}{\sqrt{\epsilon_{\pm}}}$, and  the energy of the zero modes  becomes:
\begin{eqnarray}
u&=&{1\over2}\sum_{n=1}^\infty\int{d^2k\over(2\pi)^2}
\sqrt{k^2+{n^2\pi^2\over D^2}}\left[ \frac{1}{(1 + \rho + \sigma)} + \frac{1}{ (1 + \rho - \sigma)} \right]
\nonumber \\ 
&=&\Lambda \sum_{n=1}^\infty\int{d^2k\over(2\pi)^2}
\sqrt{k^2+{n^2\pi^2\over D^2}},
\label{zeropt2}
\end{eqnarray}
where we denoted $\displaystyle \Lambda =\frac 1 2\left[ \frac{1}{(1 + \rho + \sigma)} + \frac{1}{ (1 + \rho - \sigma)} \right] $ and $k_z$ is obtained from the boundary conditions.

To evaluate Eq.~(\ref{zeropt2}) we employ dimensional regularization.
We let number the transverse dimensions be $d$, which we will 
treat as a continuous, complex variable. We also 
employ the Schwinger proper-time representation for the square root, so that we have
\begin{equation}
u=\Lambda \sum_n\int{d^dk\over(2\pi)^d}\int_0^\infty{dt\over t}t^{-1/2}
e^{-t(k^2+n^2\pi^2/D^2)}{1\over\Gamma(-{1\over2})},
\end{equation}
where we have used the Euler representation for the Gamma function.
We follow the standard procedure and carry out the Gaussian integration over $k$, 
 use the Euler representation, and  use he definition of the Riemann zeta function $\zeta$ to obtain:
\begin{equation}
u=-\Lambda{1\over2\sqrt{\pi}}{1\over(4\pi)^{d/2}}\left({\pi\over D}\right)^{d+1}
\Gamma\left(-{d+1\over2}\right)\zeta(-d-1).
\label{zeroptresult}
\end{equation}
When $d$ is an odd integer, this expression is indeterminate, but we use
$ \Gamma\left({z\over2}\right)\zeta(z)\pi^{-z/2}=\Gamma\left({1-z\over2}\right)
\zeta(1-z)\pi^{(z-1)/2}$ to rewrite (\ref{zeroptresult}) as
\begin{equation}
u=-\Lambda{1\over2^{d}\pi^{d/2+1}}{1\over D^{d+1}}\Gamma\left(1+{d\over2}\right)
\zeta(2+d).\label{scalarenergy}
\end{equation}
and obtain the final result for the energy
per unit area in the transverse direction in SME
\begin{equation}
u=-\Lambda {\pi^2\over720}{1\over D^3},
\label{casimir}
\end{equation}
where we used $\zeta(4)={\pi^4/90}$.
The force per unit area between the plates is
obtained by taking the negative derivative of $u$ with respect to $D$:
\begin{equation}
f_s=-{\partial\over\partial D}u=-\Lambda{\pi^2\over 240}{1\over D^4}.
\label{casforce}
\end{equation}
The above result represents the Casimir force due to two scalar fields of frequency $\omega_+$ and $\omega_-$. Comparing this to the
classic result of Casimir \cite{Casimir}:
\begin{equation}
f_{\rm em}=-{\pi^2\over 240}{1\over D^4}\,,
\label{casimirclassic}
\end{equation}
the deviation in SME is given by the multiplicative correction factor $\Lambda$. For  $\rho, \sigma \ll 1$, 
$\Lambda \cong 1- \rho$, so the expression is simplified to 
\begin{equation}
f_s^{\rm CPT-even} \cong (1-\rho) f_{\rm em}\,.
\end{equation}
Thus the lowest-order prediction of the SME is a reduction in the Casimir force by $\rho f_{\rm em}$.
 
 \subsubsection{Scalar Green's Function}
\label{scalargreen}

We derive the result of the previous subsection by the
Green's function approach for a scalar field.  A more general indexed Green function for the photon in the SME framework, in the Coulomb gauge,  has been given in \cite{Bailey:2004na}. We start from the equation of motion of a massless
scalar field $\phi$ produced by a source $K$ 
\begin{equation}
-(\partial_t^2\epsilon_{\pm}-{\vec \nabla}^2)\phi=K,
\label{eom}
\end{equation}
from which we deduce the equation satisfied by the corresponding
Green's function
\begin{equation}
-(\partial_t^2\epsilon_{\pm}-{\vec \nabla}^2) G(x,x')=\delta(x-x').
\label{kg}
\end{equation}
We introduce a reduced
Green's function $g(z,z')$ using the Fourier transformation
\begin{equation}
G(x,x')=\int{d^d k\over(2\pi)^d}e^{i{\vec k}\cdot({\vec x-\vec x'})}
\int{d\omega\over2\pi}e^{-i\omega(t-t')} g(z,z'),
\end{equation}
where we have suppressed the dependence of $g$ on ${\vec k}$ and $\omega$,
and have allowed $z$ on the right hand side
 to represent the coordinate perpendicular to the plates.
The reduced Green's function satisfies
\begin{equation}
\left(-{\partial^2\over\partial z^2}-\lambda_\pm^2\right)g(z,z')=\delta(z-z'),
\label{scalargreeneq}
\end{equation}
where $\lambda_\pm^2=\omega_\pm^2-k^2\equiv \omega ^2 \epsilon_{\pm}-k^2$ represent the two scalar modes.  Equation (\ref{scalargreeneq}) is to be
solved subject to the boundary conditions (\ref{eq:bc}), or
\begin{equation}
g(0,z')=g(D,z')=0.
\label{greenbc}
\end{equation}
The form of the solution, obtained by the  discontinuity method \cite{Milton:2004ya} for each of the two scalar modes, is
\begin{equation}
g(z,z')=\left\{\begin{array}{ll}
A\sin\lambda z,&0<z<z'<D,\\
B\sin\lambda(z-D),&D>z>z'>0,
\end{array}\right.
\end{equation}
where $g$ is continuous
at $z=z'$, but its derivative has a discontinuity, which is:
\begin{eqnarray}
A\sin\lambda_\pm z'-B\sin\lambda_\pm (z'-D)&=&0,\label{cont}\\
A\lambda_\pm \cos\lambda_\pm z'-B\lambda\cos\lambda(z'-D)&=&1.
\label{discont}
\end{eqnarray}
The solution to this system of equations is
\begin{eqnarray}
A&=&-{1\over\lambda_\pm}{\sin\lambda_\pm(z'-D)\over\sin\lambda_\pm D},\\
B&=&-{1\over\lambda_\pm}{\sin\lambda_\pm z'\over\sin\lambda_\pm D},
\end{eqnarray}
which gives for the reduced Green's function 
\begin{equation}
g_\pm(z,z')=-{1\over\lambda_\pm\sin\lambda_\pm D}\sin\lambda_\pm z_<\sin\lambda_\pm (z_>-D),
\label{scgreen}
\end{equation}
where $z_>$ ($z_<$) is the greater (lesser) of $z$ and $z'$.

From the Green's function we can calculate the force on the
bounding surfaces by evaluating the  stress tensor.  For a
scalar field, the stress tensor obtained from
\begin{equation}
T_{\mu\nu}=\partial_\mu\phi\partial_\nu\phi+g_{\mu\nu}{\cal L},\label{st}
\end{equation}
where the Lagrange density is
\begin{equation}
{\cal L}=-{1\over2}\partial_\lambda\phi \partial^\lambda\phi.
\end{equation}
 The normal-normal component of the
Fourier transform of the
stress tensor on the boundaries is
\begin{equation}
\langle T_{zz}\rangle={1\over2i}\partial_z\partial_{z'}g_\pm(z,z')
\bigg|_{z\to z'=0,D}={i\over2}\lambda_\pm\cot\lambda_\pm D.
\label{txx}
\end{equation}
However this expression will not give rise to a finite result  \cite{Milton:2004ya}, because we did not consider the discontinuity in $T_{zz}$. The corresponding normal-normal component of the stress tensor at $z=D$ is
\begin{equation}
\langle T_{zz}\rangle\bigg|_{z=z'=D}={1\over2i}\partial_z\partial_{z'}
g_\pm(z,z')\bigg|_{z=z'=D}={\lambda_\pm \over 2}.
\label{txxout}
\end{equation}
Thus the correct expression for the force per unit area
on the conducting surface is obtained integrating over all possible frequencies and momenta, and using complex frequencies $\omega \to i\xi,~ \lambda_\pm \to i \kappa_\pm $:
\begin{equation}
f_s=-{1\over2}\int{d^d k\over(2\pi)^d}\left [\int{d\xi\over2\pi}\kappa_+(\coth\kappa_+ D-1)+ \int{d\xi\over2\pi}\kappa_-(\coth\kappa_- D-1)\right].
\label{0tempforce}
\end{equation}
 We evaluate this integral using polar
coordinates:
\begin{equation}
f_s=-\frac 1 2 \left (\frac{1}{1+\rho+\sigma}+ \frac{1} {1+\rho - \sigma}\right){A_{d+1}\over(2\pi)^{d+1}}\int_0^\infty\kappa^d\,d\kappa{\kappa\over e^{2\kappa D}
-1}.\label{scalarint}
\end{equation}
Here $A_{d+1}$ is the surface area of a unit sphere in $d+1$ dimensions. The force per unit transverse area in the SME is
\begin{equation}
f_s = \Lambda (d+1)2^{-d-2}\pi^{-d/2-1}{\Gamma\left(1+{d\over2}\right)\zeta(d+2)\over
D^{d+2}}.
\label{scalarforce}
\end{equation}
which, for $d=2$, using $\Gamma(2) =1$ and $\zeta(4)=\frac {\pi^4} {90} $ gives 
\begin{equation}
f_s^{\rm CPT-even}=- \Lambda {\pi^2\over 240}{1\over D^4}\,,
\end{equation}
which confirms the formula obtained before, \rf{casforce}.

 \subsection{Case~II: $k_{AF} \ne 0$, Scalar Green's Function}
 If one starts from the photon part of Lagrangian \rf{EQEDlag} by setting $k_F=0$, the equation of motion is modified to:
\begin{equation}
-\partial_\alpha {F_\mu}^\alpha + (k_{AF})^\alpha \epsilon_{\mu \alpha \beta \gamma} F^{\beta \gamma} =0\,.
\end{equation}
This expression is similar to the one obtained in the analysis of the Casimir Effect in two dimensions (the so-called Maxwell-Chern-Simons Casimir Effect \cite{Milton:2004ya}).The gauge field is effectively massive, and it satisfies the equation:
\begin{equation}
\left [-\partial^2+(k_{AF})^2\right ] \epsilon^{\mu\nu\alpha\beta}\partial_\alpha A_\beta=0
\end{equation}
with the factor $k_{AF}$ playing the role of the mass. The reduced Green function
\begin{equation}
G(x,x')=\int{d^d k\over(2\pi)^d}e^{i{\vec k}\cdot({\vec x-\vec x'})}
\int{d\omega\over2\pi}e^{-i\omega(t-t')} g(z,z'),
\end{equation}
satisfies the equation: 
\begin{equation}
\left(-{\partial^2\over\partial z^2}-\lambda_{I,II}^2+ (k_{AF})2^2\right )g_{I,II}(z,z')=\delta(z-z'),
\label{scalargreeneq}
\end{equation}
where $\lambda_{I,II}^2=\omega_{I,II}^2-k^2, \, \omega_{I,II}=(1 \pm \tau)p^0$. 
This gives, for the reduced Green function:
\begin{equation}
g_{I,II}(z,z')=-{1\over\kappa_{1,2}\sin\kappa_{I,II}\, D}\sin\kappa_{1,2} z_<\sin\kappa_{I,II} (z_>-D),
\label{scgreen}
\end{equation}
where $z_>$ ($z_<$) is the greater (lesser) of $z$ and $z'$ and $\kappa_{I,II}=\lambda_{I,II}^2-(k_{AF})^2$. The Casimir force becomes (with the usual Schwinger change $\omega \to i \xi, \,\kappa \to i \rho$):
\begin{equation}
f_s=-{1\over2}\int{d^d k\over(2\pi)^d}\left [\int{d\xi \over2\pi}\rho_I(\coth\rho_I D-1)\right]-{1\over2}\int{d^d k\over(2\pi)^d}\left [\int{d\xi \over2\pi}\rho_{II}(\coth\rho_{II} D-1)\right]\,.
\end{equation}
Neglecting terms of higher order in $\sigma$, the equation can be transformed,  working in polar coordinates, into:
\begin{equation}
f_s=-{4 \pi^{\frac {(d+1)}{2}}\over \Gamma (\frac {(d+1)}{2})(2 \pi)^{(d+1)}} \int \rho^d \frac {d^d \rho \sqrt{\rho^2+(k_{AF})^2} } {e^{2a \sqrt {\rho^2+(k_{AF})^2}}-1}\,.
\end{equation}
Setting $x=2D \sqrt{\rho^2+(k_{AF})^2}, \, d=2$, and approximating, for small values of $(k_{AF}D)^2 \ll 1$, $\displaystyle (x^2-4 (k_{AF})^2D^2)^{\frac1 2} \cong x-\frac {2 (k_{AF})^2D^2}{x} $:
\begin{equation}
f_s \cong -\frac {1}{16 \pi^2 D^4} \left [\int^\infty_0 \frac{x^3 dx}{e^x-1} -2 (k_{AF})^2 D^2\int^\infty_0\frac {x dx}{e^x-1}\right ]\,, 
\end{equation}
which gives, expressed in terms of the usual Riemann zeta functions:
\begin{eqnarray}
f_s &&\cong -\frac {1}{16 \pi^2 D^4} \left [ \Gamma(4) \zeta (4)-2 (k_{AF})^2 D^2 \Gamma(2) \zeta (2)\right]\nonumber \\
&&= -\frac{\pi^2}{240 D^4}+\frac{(k_{AF})^2}{48D^2}
\end{eqnarray}
or, 
\begin{equation}
f_s^{\rm CPT-odd} =-\frac{\pi^2}{240 D^4}+\frac{(k_{AF})^2}{48D^2}\,.
\end{equation}
 One recognizes in the first term the usual scalar Casimir force, while the second represents the correction from the CPT-odd terms in the photon Lagrangian, which has and different dependence on the separation between the plates ($1/D^2$). However, it is very unlike to have any observable affect since $k_{AF}$ is constrainted to have very tiny values and appears squared in $f_s$; also the weaker separation dependence  makes the sensivity to $k_{AF}$ even weaker.
 
\section{The fermion field}
\label{fermion}

By taking the fermion part of the Lagrangian \rf{EQEDlag} and setting $e_\nu, f_\nu$ and $g_{\kappa\mu\nu}$ to zero, the modified 
Dirac equation can be obtained as:
\begin{equation}
\left \{i \left[\gamma^{\mu}+ \gamma_{\nu}(c^{\mu \nu} +d^{\mu \nu}\gamma_5)\right]D_\mu -\left[m_e+V(r)+a_\mu \gamma^\mu+b_\mu \gamma_5 \gamma^\mu  +H_{\mu \nu} \sigma^{\mu \nu}\right]    \right\} 
\psi=0,
\label{DiracS1}%
\end{equation}
with $D_\mu=i\partial_\mu+ieA_\mu$,  and we assume a conventional photon sector.

To reproduce the boundary set for the Casimir effect between two plates for massless fermions, take $V=V_0\,\theta(x)$ and $A_\mu=0$, and set $\psi(r,t)=\psi(r)e^{i\omega t}$.
\begin{equation}
\left\{ -in_i\left[\gamma^i+\left(c^{\nu i}+d^{\nu i} \gamma_5\right)\gamma_\nu\right]-1\right\}V_0\psi(r)=0
\end{equation}
where $n_\mu$ represents an outward normal at a boundary surface. Note that neither $a_\mu$, nor $b_\mu$ appears in the boundary condition.  Taking $V_0 \to\infty$, we obtain the boundary condition corresponding to the Dirac Equation for the SME:
\begin{equation}
\left\{ 1+in_i\left[ \gamma^i+\left( c^{\nu i}+d^{\nu i} \gamma_5 \right) \gamma_\nu \right] \right \} \psi(r) \bigg|_{z=0,D}=0.
\end{equation} 
Because the parameter $a_\mu$ is not physical and can be eliminated (which is equivalent to shifting the zero-point energy) and that Dirac equation is still complicated and difficult to disentangle when all parameters are swiched on at the same time, we consider the simplest non-trivial case by analyzing the effect of nonzero $b_\mu$. Under these simplifying conditions, the Dirac equation becomes:
\begin{equation}
i \gamma ^\mu \left( \partial_\mu+b_\mu \gamma_5\right) \psi(r)=0
\end{equation}
and the boundary condition
on the Dirac field $\psi$ 
\begin{equation}
(1+in\cdot \gamma)\psi\bigg|_{z=0,D}=0.
\label{bagbc}
\end{equation}
For the situation of parallel plates at $z=0$ and $z=D$, this boundary condition becomes
\begin{equation}
(1\mp i\gamma^3)\psi\bigg|_{z=0,D}=0.
\label{bcfermi}
\end{equation}

\subsection{Summing the modes}

 The easiest, but not the most rigorous method, is to sum the modes.
We introduce a Fourier transform in time and the transverse spatial
directions,
\begin{equation}
\psi(x)=\int{d\omega\over2\pi}e^{-i\omega t}\int {(d{\,\vec k\,})\over(2\pi)^2}
e^{i{\vec k\cdot \vec x}}\psi(z;{\vec k}, \omega),
\end{equation}
In what follows we choose a representation of the Dirac matrices
in which $i\gamma_5$ is diagonal. In $2\times 2$ block form, the representation for the $\gamma_5, \gamma_0$ is:
\begin{equation}
i\gamma_5=\left(\begin{array}{cc}
1&\,0\\
0&\,-1
\end{array}\right) \, ,\quad \quad 
\gamma^0=\left(\begin{array}{cc}
0&\,-i\\
i&\,0
\end{array}\right)\,,
\end{equation}
and the explicit form of all the other Dirac matrices follow
from
$\mbox{ $\vec \gamma$ }=i\gamma^0\gamma_5\mbox{$\vec\sigma$}$.
We obtain two equations (taking for simplicity )
\begin{eqnarray}
\left[ (\omega +ib_z\pm \left( ib_0+i{\partial\over\partial z}\right) \right ] u_\pm+\left(i b_+ \mp  k_+ \right) v_\pm&=&0,
\label{deqn1}\\
 \left(ib_- \mp k_- \right) u_\pm+\left[ \omega-ib_z \pm \left(-i{\partial\over\partial z}+ib_0)\right)\right]v_\pm&=&0\,,
\label{deqn2}
\end{eqnarray}
where the subscripts indicate the eigenvalues of $i\gamma_5$
and $u$ and $v$ are eigenvectors of $\sigma^3$ with eigenvalue $+1$ or $-1$,
respectively. In the above, we used the notation $k_\pm=k_x \pm ik_y, \, b_\pm=b_x \pm ib_y$. The equations simplify considerable if we take ${\vec k}$  along the  $y$ axis and $b=(0,b_x,0,0)$, (or similarly, $\vec k$ along the x-axis, and $\vec b$ along the y axis) and we obtain:
\begin{equation}
\left(\lambda^2+{\partial^2\over\partial z^2}\right) \psi=0\,,
\end{equation}
where $\lambda^2=\omega^2-k^2+b_x^2$. From the boundary conditions
\begin{equation}
\lambda D =\left( n+1/2\right)\pi\,, \quad \omega=\sqrt{k^2-b_x^2+\left( n+1/2 \right)^2\frac{\pi^2}{D^2}}\,.
\end{equation}

When we compute the zero-point energy, we must sum over
odd integers, remember that there are two modes, and taking into account the
minus sign (for fermion energy):
\begin{eqnarray}
u&=& -2{1\over2}\sum_{n=0}^\infty \int {(d{\,\vec k\,})\over(2\pi)^2}
\sqrt{k^2-b_x^2+{(n+1/2)^2\frac{\pi^2}{D^2}}}\nonumber\\
&=&{1\over2\sqrt{\pi}}{1\over4\pi}\sum_{n=0}^\infty\int_0^\infty {dt\over t}
t^{-3/2}e^{-t\left[-b_x^2+(n+1/2)^2\pi^2/D^2\right]}\nonumber\\
&=&{1\over8\pi^{3/2}}\Gamma\left(-{3\over2}\right)\sum_{n=0}^\infty
{\left(\left(n+\frac{1}{2}\right)^2\frac{\pi^2}{D^2}-b_x^2\right)^{3\over 2}}\,.
\end{eqnarray}
Using $\frac 1 8 \sum_{n=0}^\infty \left( 2n+1\right)^3=-\frac 7 8 \zeta(-3)$
 and $\zeta(-3)=-B_4/4=1/120$, we obtain the Casimir energy for a fermion in the SME:
\begin{equation}
u=\frac 7 8  \frac {\pi^2} {720D^3}+\frac {1}{48} \frac {b_x^2}{D}
\end{equation} 
and correspondingly, the Casimir force 
 \begin{equation}
f_f=-{\partial\over\partial D} u=-\frac 7 8  \frac {\pi^2} {240D^4}-\frac {1}{48} \frac {b_x^2}{D^2}\,.
\label{smefermion}
\end{equation} 
The first term in this expression represents the conventional fermion Casimir force. The second term represents the shift due to the Lorentz-violating terms in SME.

\subsection{Green's Function Method}

A more controlled calculation starts from the equation
satisfied by the Dirac Green's function,
\begin{equation}
\gamma_\mu (\partial^\mu+b^\mu \gamma_5) G(x,x')=i\delta(x-x'),
\end{equation}
subject to the boundary condition
\begin{equation}
(1+i{\vec n\cdot \mbox{$\vec\gamma$}})G\bigg|_{z=0,D}=0.
\end{equation}
We introduce a reduced, Fourier-transformed, Green's function,
\begin{equation}
G(x,x')=\int{d\omega\over2\pi}e^{-i\omega(t-t')}\int{(d{\,\vec k\,})\over(2\pi)^2}
e^{i{\vec k\cdot(\vec x-\vec x')}}g(z,z';{\vec k}, \omega),
\end{equation}
which satisfies
\begin{equation}
\left [ -\gamma^0 \omega+ {\vec\gamma}\cdot \left ({\vec k}+{\vec b} \gamma_5 \right )-i\gamma^3 
\displaystyle{\partial\over \partial z} \right ] g(z,z')=\delta(z-z').
\end{equation}
 Introducing  the components of $g$ corresponding to the $+1$ or $-1$
eigenvalues of $i\gamma_5$,
\begin{equation}
g=\left(\begin{array}{cc}
g_{++}&g_{+-}\\
g_{-+}&g_{--}
\end{array}\right),
\end{equation}
we insert these Green's functions into the vacuum expectation value
of the energy-momentum tensor.  The latter is
\begin{equation}
T^{\mu\nu}={1\over4i}\psi\gamma^0\left(\gamma^\mu\partial^\nu
+\gamma^\nu\partial^\mu\right)\psi+g^{\mu\nu}{\cal L},
\end{equation}
where ${\cal L}$ is the fermionic part of ${\cal L}^{QED}$ in \rf{EQEDlag} with only $b_\mu$ nonzero.
We take the vacuum expectation value by the usual replacement $\psi\psi\gamma^0\to -iG$,
 where $G$ is the fermionic Green's function, and obtain
\begin{equation}
\langle T_{zz}\rangle=i\lambda\tan\lambda D,
\end{equation}
where $\lambda^2=\omega^2-k^2+b_x^2 \equiv -\xi^2-k^2+b_x^2$. The force per unit area is
\begin{eqnarray}
f_f&=&\int{d^2k\over(2\pi)^2}\int{d\omega\over2\pi} i\lambda\tan\lambda D\,.
\end{eqnarray}
Introducing polar coordinates $k= \kappa \sin \theta$ and $\left (\xi^2-b_x^2\right)^{1 \over 2}= \kappa \cos \theta$,
\begin{eqnarray}
\displaystyle
dk d \xi &=&\frac {\kappa^2 \cos \theta}{\xi} d \kappa d \theta=\frac {\kappa d \kappa d \theta} {\left(1 +\displaystyle \frac{b_x^2}{\kappa^2 \cos^2 \theta}\right)^{\frac 1 2}}
\nonumber \\
&\cong & \kappa d\kappa d \theta \left (1-\frac 1 2 \frac{b_x^2}{\kappa^2 \cos^2 \theta} \right)\,,
\end{eqnarray}
we obtain:
\begin{eqnarray}
f_f&=&\int{d^2k\over (2\pi)^2}\int{d\xi\over2\pi}\kappa  \tanh\kappa D\nonumber
\\
&=&{1\over 2\pi^2}\int_0^\infty d \theta d\kappa\,\kappa^3\left[1-{2\over e^{2\kappa D}+1}\right]\left (1-\frac 1 2 \frac{b_x^2}{\kappa^2 \cos^2 \theta} \right)\,.
\end{eqnarray}
As in \cite{Milton:2004ya}, we omit the first term in the last square bracket, as the same term is present
in the vacuum energy outside the plates. Using
\begin{equation}
\int_0^\infty{x^{s-1}\,dx\over e^x+1}=(1-2^{1-s})\zeta(s)\Gamma(s),
\end{equation}
we find
\begin{equation}
f_f=-\frac 7 8 \frac {2A_3}{(2 \pi)^3}\frac{1}{(2D)^3}\zeta(4) \Gamma(4)-\frac{b_x^2(4 \pi)}{(2 \pi)^3}\frac{1}{(2D)^2}\zeta(2) \Gamma(2)
\end{equation}
with $A_3$ the area of the unit sphere in 3 dimensions; which is, indeed, 
 \begin{equation}
f_f=-{\partial\over\partial D} u=-\frac 7 8  \frac {\pi^2} {240D^4}-\frac {1}{48} \frac {b_x^2}{D^2}\,,
\end{equation} 
the same as \rf{smefermion}.
 

\section{Summary and Conclusion}
\label{summary}

We have presented an analysis of the Casimir force in the extended QED derived from the minimal version of the Standard Model (SME) which includes Lorentz-violating terms, for both the scalar  and fermion fields. 
Thinking of the two polarizations of the photon as two scalar fields, propagating with different frequencies, we calculated  the leading order of the deviation from the classical expression for the Casimir force for two parallel conducting plates. In the case where the Lorentz violation is induced by CPT-even effects only, the correction term on the Casimir force is multiplicative and of the form $(1- \rho)f_{\rm em}$, with $f_{\rm em}$ the conventional Casimir force and $\rho$ a small parameter caracterizing the Lorentz violation. Since the Casimir force is measured to only 15\% precision, no useful bounds on the $\rho$ parameter can be obtained from such an expression. In the case in which Lorentz violation is induced by the CPT-odd terms in the Lagrangian, the effective Lagrangian gives rise to an
additive term as correction; additionally this term has a different dependence on the distance $D$ between the plates from the usual Casimir force: $1/D^2$ versus $1/D^4$. Unfortunately in this case  the bounds obtained on the $k_{AF}$ parameter are too weak to have an experimental significance at this time. For the case of the fermion field, the correction term in the SME is also additive, and also of the form $b^2 / D^2$, where $b$ is measure of Lorentz violation. In all cases, the leading order contribution is quadratic in Lorentz violating parameters, which makes the experimental sensitivity much weaker. 

In conclusion, we calculated the deviation of the Casimir force in the Lorentz-violating extension of the standard model from its standard quantum electrodynamics value. We have shown that, while small, it does not contradict any experimental measurements, unlike in some theories with extra dimensions \cite{Linares:2005cj}. The deviation predicted is of theoretical interest, and would only be useful in setting any significant constraints on the parameters of the model only if the  precision of the experimental measurements will increase significantly.

\begin{acknowledgments}
 We thank NSERC of Canada for partial financial support (SAP0105354). We are also grateful to V. Alan Kosteleck\'y for his comments and suggestions.
  \end{acknowledgments}



\end{document}